\documentstyle[12pt,epsf]{article}
\begin{document}
\baselineskip 16pt plus 2pt minus 2pt

\newcommand{\beq}{\begin{equation}}
\newcommand{\eeq}{\end{equation}}
\newcommand{\beqa}{\begin{eqnarray}}
\newcommand{\eeqa}{\end{eqnarray}}
\newcommand{\dida}[1]{/ \!\!\! #1}
\renewcommand{\Im}{\mbox{\sl{Im}}}
\renewcommand{\Re}{\mbox{\sl{Re}}}
\def\simge{\hspace*{0.2em}\raisebox{0.5ex}{$>$}
     \hspace{-0.8em}\raisebox{-0.3em}{$\sim$}\hspace*{0.2em}}
\def\simle{\hspace*{0.2em}\raisebox{0.5ex}{$<$}
     \hspace{-0.8em}\raisebox{-0.3em}{$\sim$}\hspace*{0.2em}}

\begin{titlepage}


\hfill{TRI-PP-99-01}

\vspace{1.0cm}

\begin{center}
{\large {\bf Rare Pionium Decays and Pion Polarizability}}

\vspace{1.2cm}

H.-W. Hammer\footnote{email: hammer@triumf.ca} and
J.N. Ng\footnote{email: misery@triumf.ca}

\vspace{0.8cm}

TRIUMF, 4004 Wesbrook Mall, Vancouver, B.C., Canada V6T 2A3\\[0.4cm]
\end{center}

\vspace{1cm}

\begin{abstract}
We calculate the decay of pionium atoms into two photons.
The pion polarizabilities give rise to a 10\% correction to 
the corresponding decay width for pointlike pions. This opens the
possibility to obtain the difference between the electric and 
magnetic polarizability of the charged pion from a future 
measurement of the branching fraction of pionium into two photons.
For such an experiment the $\pi\pi$-scattering
lengths would have to be known to better than $5\%$ precision.
We also comment on the contribution of the axial anomaly
to the decay of pionium into $\gamma\pi^0$.
\\[0.3cm]
\end{abstract}

\vspace{2cm}
\vfill
\end{titlepage}

The response of a composite particle to an external electromagnetic 
field is described by its electric ($\alpha$) and magnetic ($\beta$)
polarizabilities \cite{Hol90}. 
They are fundamental quantities whose understanding
is of great importance in any model or theory of the strong
interaction. In particular, the polarizabilities of the pion allow
for a clean test of Chiral Perturbation Theory (ChPT). To one loop order
only two of the ten low energy constants at ${\cal O}(p^4)$ contribute
and ChPT predicts $\alpha_{\pi^\pm}=-\beta_{\pi^\pm} = 2.8 \cdot
10^{-4} \mbox{ fm}^3$ \cite{Hol90}.\footnote{In the following, we will
express the polarizabilities in units of $10^{-4} \mbox{ fm}^3$.}
To this order one has
$\alpha_\pi=-\beta_\pi$ because the lowest order coupling to the 
electromagnetic field is proportional to $F_{\mu\nu} F^{\mu\nu} \sim 
(\vec{E}^2-\vec{B}^2)$. Since higher order corrections are suppressed
this relation is expected to hold approximately to all orders. More 
recently, B{\"u}rgi performed a two loop calculation and obtained
$\alpha_{\pi^\pm}=2.4 \pm 0.5$ and $\beta_{\pi^\pm}=-2.1 \pm 0.5$ 
\cite{Bue96}. 

Experimentally both  of these quantites are measured using Compton 
scattering off pions. However, since a pion target is not available, the 
process is only indirectly accessible when embedded in other reactions.
As a consequence, experiments are difficult and 
the polarizabilities are relatively poorly known. For example one of 
the experiments done is to use radiative pion scattering off a heavy nucleus, 
$\pi^- Z \to \pi^- Z \gamma$. In this case the pion scatters off a virtual 
photon in the Coulomb field of the nucleus (Primakoff scattering).
This method was used in the Serpukhov experiment \cite{Ant83},
resulting in $\alpha_{\pi^\pm}=-\beta_{\pi^\pm}= 6.8\pm 1.4 \pm 1.2$.
A similar way is to use the radiative pion photoproduction process,
$\gamma p \to  \gamma \pi^+ n$, where the incoming photon scatters 
off a virtual pion. Such an experiment was performed at
the Lebedev Institute \cite{Aib86} with the result
$\alpha_{\pi^\pm}=-\beta_{\pi^\pm}=20 \pm 12$.
Both experiments are in variance with the results from ChPT 
\cite{Hol90,Bue96}. Furthermore, they suffer from a limited statistical 
accuracy and systematic errors. The extrapolation from the actual data to 
the Compton scattering amplitude in the case of radiative pion 
photoproduction has recently been studied in Ref. \cite{DrF94} 
and a new experiment has been proposed at Mainz. 
An overview of proposed Primakoff experiments is given in Ref. \cite{MoS98}.

In a second type of experiment, one makes use of the crossed channel 
reaction $\gamma\gamma \to \pi\pi$ \cite{cross,HoD93} and attempts
to extract the polarizabilities from the cross-section.
This reaction also has the advantage of being able to determine the 
neutral pion polarizability as well. However, it 
suffers from the difficulties of having to deal with
strong final state interaction. 
The most recent analysis \cite{HoD93} is consistent with the result 
from ChPT but the experimental error bars can accomodate differences
in polarizability calculations by a factor of two. 

In light of the present discrepancy
between theory and experiment, we propose to study the decay mode of
pionium into two photons. The formation of pionium atoms has been observed
recently \cite{Afa94}. They  decay in more than $99\%$ of the time  
into two neutral pions. Furthermore, pionium can decay into two photons
and $\gamma\pi^0$ as well. The DIRAC Collaboration at CERN aims to
measure the pionium lifetime with $10\%$ precision in order to
extract the $\pi\pi$-scattering lengths \cite{Sch98}. 
Recently, the decay of pionium atoms has been studied using 
nonrelativistic effective field theories \cite{Hol99} and a general 
expression that allows the extraction of the $\pi\pi$-scattering lengths
has been given in this framework as well \cite{GaG99}. In this note
we point out that if the next generation of pionium experiments can
measure the branching ratio $\Gamma(\mbox{pionium}\to \gamma\gamma)/
\Gamma_{tot}$ sufficiently accurate, it should be able 
to provide a clean determination of the charged
pion polarizability. Below, we will calculate the decay pionium$\to\gamma
\gamma$. The theoretical assumptions required are crossing symmetry
and the usual formulation of bound states in quantum field theory.
We  show that the pion polarizability
provides corrections of the order of $10\%$ to the point particle result
for $\Gamma(\mbox{pionium}\to \gamma\gamma)$.

We first consider Compton scattering off charged pions,
\beq
\gamma(q_1,\epsilon_1)+\pi^+(p_1)\to\gamma(q_2,\epsilon_2)+\pi^+(p_2)\;.
\eeq
The polarizabilities are defined by the threshold expansion of
the Compton scattering amplitude in the laboratory frame,
\beq
{\cal M}_{\gamma\pi\to\gamma\pi}=8\pi m_\pi \left[ \left(
-\frac{\alpha}{m_\pi} + \omega_1 \omega_2 \alpha_\pi \right) 
\vec{\epsilon_1}\cdot \vec{\epsilon_2} +(\vec{\epsilon_1} \times 
\vec{q_1})\cdot (\vec{\epsilon_2} \times \vec{q_2}) \beta_\pi \right]\;,
\eeq
where $\alpha=e^2/4\pi$ is the electromagnetic fine structure constant and
$m_\pi$ is the charged pion mass.
$\omega_{1(2)}$, $\vec{q}_{1(2)}$, and $\vec{\epsilon}_{1(2)}$ are the 
energy, momentum, and polarization vectors of the photon in the initial
(final) state, respectively.
The Compton tensor $M_{\mu\nu}$ is defined by
\beq
{\cal M}= i e^2 M_{\mu\nu} \epsilon_1^\mu \epsilon_2^\nu\;.
\eeq
The structure of $M_{\mu\nu}$ is given by \cite{Ter73,DoH89,DHW93},
\beqa
\label{comten}
M_{\mu\nu} &=&\int d^4 x \,e^{i q_2 \cdot x} \langle \pi^+ 
(p_2) | T[ J_\nu^{EM}(x) J_\mu^{EM}(0)]| \pi^+ (p_1)\rangle\\
&=& -\frac{T_\mu(p_1, p_1+q_1)T_\nu(p_2 +q_2, p_2)}
     {(p_1+q_1)^2 -m_\pi^2}-\frac{T_\mu(p_2-q_1, p_2)T_\nu(p_1, p_1-q_2)}
     {(p_1-q_2)^2 -m_\pi^2}
\nonumber\\
& &+2 g_{\mu\nu} +\frac{1}{3}\langle r_\pi^2\rangle (q_1^2 g_{\mu\nu}
-q_{1\mu} q_{1\nu} +q_2^2 g_{\mu\nu} -q_{2\mu} q_{2\nu}) \nonumber\\
& &-\frac{m_\pi}{\alpha}(\alpha_\pi-\beta_\pi)(q_1 \cdot q_2 g_{\mu\nu}
-q_{2\mu} q_{1\nu}) +\ldots \nonumber\;,
\eeqa
where
\beq
\label{pivert}
T_\mu(p_i, p_f) = (p_i+p_f)_\mu \left[ 1+\frac{1}{6}\langle r_\pi^2\rangle
q^2 \right]+q_\mu \frac{1}{6}\langle r_\pi^2\rangle (p_i^2-p_f^2) +\ldots\;,
\eeq
is the electromagnetic vertex of the pion and $q=p_f-p_i$.
The dots stand for higher order loop corrections which
we will neglect in the following. The polarizabilities
appear only in the combination $\alpha_\pi-\beta_\pi$ since to
this order the coupling to the electromagnetic field is proportional
to $F_{\mu\nu} F^{\mu\nu} \sim (\vec{E}^2-\vec{B}^2)$.

For the pionium decay into two photons, we need the crossed
amplitude $\tilde{M}_{\mu\nu}$ for the reaction
\beq
\pi^+(p_1) +\pi^-(p_2)\to\gamma(q_1,\epsilon_1)+\gamma(q_2,\epsilon_2)\;,
\eeq
which is readily obtained from Eq. (\ref{comten}) by the substitutions 
$(p_2 \to -p_2)$ and $(q_1 \to -q_1)$,
\beqa
\label{pionten}
\tilde{M}_{\mu\nu} &=& -\frac{T_\mu(p_1, p_1-q_1)T_\nu(q_2 -p_2, -p_2)}
     {(p_1-q_1)^2 -m_\pi^2}-\frac{T_\mu(q_1-p_2, -p_2)T_\nu(p_1, p_1-q_2)}
     {(p_1-q_2)^2 -m_\pi^2} \\
& &+2 g_{\mu\nu} +\frac{1}{3}\langle r_\pi^2\rangle (q_1^2 g_{\mu\nu}
-q_{1\mu} q_{1\nu} +q_2^2 g_{\mu\nu} -q_{2\mu} q_{2\nu}) \nonumber\\
& &+\frac{m_\pi}{\alpha}(\alpha_\pi-\beta_\pi)(q_1 \cdot q_2 g_{\mu\nu}
-q_{2\mu} q_{1\nu}) +\ldots \nonumber\;.
\eeqa
Using Eq. (\ref{pionten}), we calculate the width for the decay
pionium$\to \gamma\gamma$ at leading order. The decay width $\Gamma$ 
is given by
\beq
\label{width}
d\Gamma = |\psi(0)|^2 \frac{1}{4 m_\pi^2}\frac{1}{2}|{\cal M}|^2
\frac{d^3 q_1}{2 q_{10} (2\pi)^3}\frac{d^3 q_2}{2 q_{20} (2\pi)^3}
(2\pi)^4 \delta^4(p_1+ p_2 -q_1 -q_2)\;,
\eeq
where $\psi(0)$ is the Coulomb wave function for the $\pi^+ \pi^-$
bound state at the origin.
Evaluating Eq. (\ref{width}) in the pionium rest frame,
the pion charge radius does not contribute and we obtain
\beq
\label{eq9}
\Gamma = \frac{2\pi\alpha^2}{m_\pi^2}|\psi(0)|^2 \left[1+
\frac{m_\pi^3}{\alpha}(\alpha_\pi-\beta_\pi)\right]^2\;.
\eeq
From the Coulomb wave function, we have  
\beq
|\psi(0)|^2=\frac{1}{\pi a_B^3}=\frac{1}{\pi}\left( \frac{m_\pi \alpha}{2}
\right)\;,
\eeq
and finally obtain
\beq
\label{decgg}
\Gamma(\mbox{pionium}\to \gamma\gamma) = \frac{m_\pi \alpha^5}{4}\left[1+
\frac{m_\pi^3}{\alpha}(\alpha_\pi-\beta_\pi)\right]^2\;.
\eeq
The corresponding formula without the polarizability corrections
has already been given in Ref. \cite{UrP61}. Using the central
value of $\alpha_{\pi^\pm}=
-\beta_{\pi^\pm}= 6.8\pm 1.4 \pm 1.2$ \cite{Ant83}, we find
\beq
\label{decggnum}
\Gamma(\mbox{pionium}\to \gamma\gamma) = 0.723\, \mbox{meV}\cdot
\left[1+0.132 +0.004\right]\;.
\eeq
The correction of the polarizability to the pointlike result is
of the order $10\%$ which allows for a clean measurement of the
pion polarizability.\footnote{A similar sensitivity to the polarizabilities 
is observed in the reaction $\gamma\gamma\to \pi^+ \pi^-$ (see e.g.
Ref. \protect\cite{Bue96}). The pionium decay, however, has the advantage that
the pions annihilate at rest and we obtain the amplitude directly at
threshold.}
By far the dominating decay of the pionium, however, is the one into
two neutral pions. The corresponding decay width 
is given by \cite{Des54} (see also Ref. \cite{Nem85}),
\beq
\label{deccpp}
\Gamma(\mbox{pionium} \to \pi^0 \pi^0)=|\psi(0)|^2\frac{16\pi}{9}
(a_0^0 -a_0^2)^2 \sqrt{\frac{\Delta m_\pi}{m_\pi}}\;,
\eeq
where $a_0^0$ $(a_0^2)$ is the $s$-wave $\pi\pi$-scattering length
for isospin 0 (2) and $\Delta m_\pi=m_\pi-m_{\pi^0}$.
Approximating $\Gamma_{tot}\approx \Gamma(\mbox{pionium} \to \pi^0 \pi^0)$, 
we obtain the branching ratio
\beq
\label{branch}
\frac{\Gamma(\mbox{pionium} \to \gamma\gamma)}{\Gamma_{tot}}=
\left\{\frac{9 \alpha^2}{8}\left[ m_\pi (a_0^0 -a_0^2)\right]^{-2}
\left(\frac{\Delta m_\pi}{m_\pi}\right)^{-1/2} \right\}\left[1+
\frac{m_\pi^3}{\alpha}(\alpha_\pi-\beta_\pi)\right]^2\;,
\eeq
where the factor in the curly bracket is about $0.28\%$.
As seen in Eq. (\ref{branch}), the pionium wave function has dropped out 
and the associated uncertainties are eliminated. The main source 
of uncertainty is now the value of the $\pi\pi$-scattering
lengths. The current experimental value for $(a_0^0 -a_0^2)
=(0.288\pm 0.051) m_\pi^{-1}$ is taken from Ref. \cite{Dum83}. 
In order to be able to extract the pion polarizabilities,
the uncertainty introduced by the $\pi\pi$-scattering lengths has 
to be smaller than the $10\%$ correction the 
polarizabilities give to the pointlike result.
Consequently, one needs to know $(a_0^0 -a_0^2)$ with a $5\%$
precision or better which appears to be an attainable goal
for the future. When the smaller value $\alpha_{\pi^\pm}-
\beta_{\pi^\pm}= 4.5\pm 1.0$ \cite{Bue96} predicted by ChPT
is used, $(a_0^0 -a_0^2)$ has to be known to higher precision.

We also note that for the actual extraction of $\alpha_{\pi^\pm}-
\beta_{\pi^\pm}$ from experiment, the systematic inclusion of 
radiative corrections in Eq. (\ref{branch}) is needed.
If, e.g.,  the one-loop amplitude for $\pi^+ \pi^- \to \gamma\gamma$
(see Ref. \cite{Bue96}) instead of the tree-level 
result is used, we obtain a correction of 0.042 to the 1 in the 
square bracket of Eqs. (\ref{eq9}, \ref{decgg}, \ref{decggnum}, 
\ref{branch}). This correction does not affect the sensitivity 
to $\alpha_{\pi^\pm}-\beta_{\pi^\pm}$. However, it is not
negligible compared to the effect of the polarizabilities
and therefore needed for a sensible extraction from experiment. 
Furthermore, there are corrections from the strong interaction
to the pionium wave function which cancels in Eq. (\ref{branch})
to leading order (See e.g. Ref. \cite{GaG99} and references therein).
However, since the suggested measurement is beyond the present experimental
capabilities, such a higher order calculation would be premature.

We next turn to the decay of pionium into $\gamma\pi^0$. This
branch is to leading order determined by the
axial anomaly of QCD. In principle this opens the possibility of
measuring the anomaly via the decay pionium$\to\gamma\pi^0$ which,
however, is outside the scope of this note.
The amplitude for $\pi^+ (p_1)+ \pi^- (p_2) \to \gamma (q_1, \epsilon_1) 
+ \pi^0 (q_2)$ is given by \cite{WZW71,Hol96}
\beq
\label{anom}
{\cal M} = \frac{e}{4\pi^2 F_\pi^3}\epsilon^{\mu\nu\alpha\beta}
\epsilon_{1\mu} q_{2\nu} p_{1\alpha} p_{2\beta}\;.
\eeq
Naively one would expect this to be order $\alpha$ of the dominant two neutral
pion decays. A close examination of Eq. (\ref{anom}) reveals that this mode 
is suppressed by the relative three momentum of the initial state pions. 
Since at threshold where pionium is formed $p_1=p_2=(m_\pi, \vec{0})$,
this contribution
vanishes and the anomaly contributes only at ${\cal O}(\vec{p})$.
Moreover, this result justifies our approximation of the total
width by $\Gamma(\mbox{pionium}\to  \pi^0 \pi^0)$ in Eq. (\ref{branch}).

In  conclusion, we have presented the calculation of the contribution of
pion polarizablities to the branching ratio of the rare two photons decays
of pionium. As seen in Eq. (\ref{branch}), the only uncertainty 
involves the values of the $\pi\pi$-scattering 
lengths which one can determine accurately. In fact, these scattering
lengths will be measured by a first generation pionium experiment
at CERN \cite{Sch98}. If the scattering
lengths can be determined to better than $5\%$ accuracy, the branching
ratio into two photons offers a clean way of determining the charged pion
polarizabilites. The remaining theoretical uncertainties involve
higher order loop effects and are expected to be small. In addition, this
measures $\alpha_{\pi^\pm}-\beta_{\pi^\pm}$ at threshold which is accessible
currently only by extrapolation and hence plagued with uncertainties. 
We are aware that rare decays of pionium
are dauntingly difficult experiments. However, in view of the importance
of polarizabilities in hadronic physics and the clean nature of the decay
we believe it is worthwhile pursuing it vigorously.

\section*{Acknowledgement}
This work has been supported by the Natural Sciences
and Engineering Research Council of Canada.

\end{document}